\documentclass[aps,physrev,reprint,superscriptaddress]{revtex4-2}

\usepackage{amsmath} 
\usepackage{amssymb}
\usepackage{graphicx} 
\usepackage{float}
\usepackage{helvet}
\usepackage{makecell}
\usepackage{multirow}
\usepackage{siunitx}
\usepackage{booktabs} 
\usepackage{makecell} 
\usepackage[colorlinks=true,linkcolor=blue,filecolor=magenta,
urlcolor=blue]{hyperref}

\usepackage{xr}       
\usepackage{threeparttable}
\renewcommand{\arraystretch}{1.25}

\usepackage{soul}
\usepackage[dvipsnames]{xcolor}
\usepackage[version=4]{mhchem}

\usepackage{algorithm}
\usepackage{algpseudocode}
\usepackage{caption}

\begin{document}

\title{Solver-Agnostic Implementation of Atom-Informed Thermal Conductivity Fields in Continuum Heat-Flow Simulations}

\author{W. Downs}
\affiliation{Department of Mechanical Engineering, Center for Advanced Materials Processing, Ohio University, Athens, OH, USA}

\author{C. Ugwumadu}
\email{cugwumadu@lanl.gov}
\affiliation{Quantum \& Condensed Matter (T-4) Group, Los Alamos National Laboratory, Los Alamos, NM, USA}

\author{M. Ali}
\affiliation{Department of Mechanical Engineering, Center for Advanced Materials Processing, Ohio University, Athens, OH, USA}

\author{R. M. Tutchton}
\affiliation{Quantum \& Condensed Matter (T-4) Group, Los Alamos National Laboratory, Los Alamos, NM, USA}

\begin{abstract}
A recent work introduced the Simulator Collection for Atomic-to-Continuum Scales (SCACS) toolkit, a framework for improving finite element predictions of heat flow by mapping atom-resolved thermal conductivity into the stiffness matrix of the Galerkin finite element formulation [Ugwumadu \textit{et al.}, Phys. Rev. Materials 10, 053804 (2026)]. Here, we demonstrate that SCACS-derived conductivity fields are solver-independent and can be transferred to existing continuum simulation platforms. As a proof of concept, we map SCACS-derived conductivity fields from complex silicon structures onto finite element meshes in Abaqus\textsuperscript{\textregistered} and compare the resulting heat-flow solutions with that obtained using conventional uniform-conductivity assignment within Abaqus. Comparison of the two implementations shows that atom-informed conductivity fields can be incorporated into existing finite element workflows and improve realistic prediction and the accuracy of its solution. This work supports broader efforts to improve the predictive capability of continuum simulations for efficient materials design and property prediction.
\end{abstract}
\maketitle

\section{Introduction}

Continuum modeling remains one of the most practical simulation approaches for connecting materials theory to engineering-scale experiments and device-level predictions. In heat-flow simulations, the accuracy of the continuum solution depends strongly on the thermal conductivity field assigned to the material. For a material with thermal conductivity tensor \(\overleftrightarrow{K}(\mathbf{x})\) and volumetric heat-generation rate $f(x)$ at position $x$, steady-state energy balance requires \cite{zienkiewicz2005finite}
\begin{equation}\label{eq:continuityRelation}
\nabla^\mathsf{T}\overrightarrow{q}(\mathbf{x}) + f(x) = 0 ,
\end{equation}
with the heat-flux given by Fourier's law,
\begin{equation}\label{eq:Fourier}
\overrightarrow{q}(\mathbf{x}) = -\overleftrightarrow{K}(\mathbf{x})\nabla T(\mathbf{x}) .
\end{equation}
Thus, the quality of the predicted temperature and heat-flux fields depends directly on how accurately \(\overleftrightarrow{K}(\mathbf{x})\) represents the underlying material. 

In many practical finite-element workflows, however, the conductivity is treated as a uniform, experimentally fitted, or otherwise simplified quantity \cite{Kaminski2021Homo,Basaula2022,Meng2025,Yang2025}. This approximation can obscure spatial variations, anisotropy, and microstructure-dependent transport pathways that are essential for realistic heat-flow prediction in heterogeneous materials \cite{Ugwumadu2026SCACS}. This limitation becomes especially important for complex materials where local structure strongly affects transport. Examples include materials for chip design, spacecraft and plasma-facing materials.

In finite-element modeling, the material domain is discretized into subvolumes \(\Omega\) and boundary surfaces \(\Gamma\), and the assigned element-wise properties enter directly into the matrix-vector form of the Galerkin heat-flow problem \cite{zienkiewicz2005finite,Ugwumadu2026SCACS}. If the conductivity field is poorly resolved or contains artificial discontinuities, the resulting heat-flux can be inaccurate, leading to poor energy balance, mesh-dependent artifacts, or unstable solutions \cite{Powers2004}. A physically informed and smoothly varying conductivity field is therefore needed to improve both the accuracy and robustness of continuum heat-flow simulations.

We recently addressed this challenge by introducing the Simulator Collection for Atomic-to-Continuum Scales (SCACS) toolkit \cite{Ugwumadu2026SCACS}, a framework that coarse-grains atom-resolved, microstructure-aware thermal conductivity into a continuum conductivity field. This field can be mapped directly into the finite-element stiffness matrix for heat flow calculations. 

In SCACS, conductive pathways are first learned from atomistic simulations using the site-projected thermal conductivity (SPTC) method \cite{Ugwumadu2025SPTC,Gautam2025SPTC,Nepal2025GST, UgwumaduPSSB2025} on system sizes with $\sim10^3$ atoms. A graph neural network is then used to extend these predictions to large atomic models containing \(\gtrsim 10^5\) atoms, which are sufficiently large to serve as representative volume elements (RVE) for finite-element analysis \cite{Cheng2021,Ochs2026,Ugwumadu2024Cfoam,Geers2010}. The resulting atom-informed conductivity is mapped into the continuum volume using Gaussian broadening, producing an intrinsic conductivity field without \textit{ad hoc} fitting or manual assignment of matrix-element properties. Related atomic-to-continuum transfer ideas have also been used to predict mechanical stress--strain behavior in porous carbon foams through dimension-invariant scaling (fractals) \cite{Ugwumadu2024Cfoam, Uguwmadu2024NPC}.

Following the development of SCACS, the next step is to demonstrate that SCACS-derived conductivity fields are \emph{solver-independent} and can therefore be transferred to existing continuum modeling frameworks. In this work, we develop a protocol for transferring SCACS-based mesh information into Abaqus\textsuperscript{\textregistered} and solving the corresponding heat-flow problem within its finite-element environment. The caveat for the transferability of SCACS-derived fields into any solver is that the solver accepts point-cloud data, which is a widely used 3D object format \cite{Liu2021}.

The contribution of this work is three-fold. First, we show that SCACS can be integrated into existing continuum workflows, allowing users to retain their preferred finite-element platform while incorporating spatially-varying material fields. Second, we demonstrate that SCACS-derived conductivity fields improve heat-flow predictions compared with conventional approaches based on uniform or experimentally fitted conductivity values. Third, we address how direct mapping of these microstructure-aware fields have the potential to improve the efficiency and physical consistency of the mesh-refinement process. We also provide all the structural models and a \texttt{python} utility file developed in this work, as supporting materials \cite{Zenodo}.

\section{Models and Methods}\label{sec:methods}

This work focuses on transferring SCACS-derived conductivity fields into Abaqus-compatible meshes and analyzing the resulting heat-flow solutions. Complete details of the SPTC method used within SCACS are provided in Ref.~\cite{Ugwumadu2025SPTC}. The coarse-graining procedure used to convert atom-resolved conductivity into finite-element conductivity fields is described in Ref.~\cite{Ugwumadu2026SCACS}.

SCACS outputs spatially resolved orthotropic thermal conductivity fields in \texttt{.xdmf/.hdf5} format as point-cloud data. The extracted conductivity components are \(K_{xx}\), \(K_{yy}\), and \(K_{zz}\), corresponding to the \(x\), \(y\), and \(z\) directions, respectively. A representation of the \(K_{xx}\) input format is shown in Table~\ref{tab:Kxx_nodal_data_analytic_field}. These nodal fields were extracted using ParaView \cite{ParaView}. 

\begin{table}[h!]
\scriptsize
\setlength{\tabcolsep}{12pt}
\renewcommand{\arraystretch}{1.1}
\centering
\caption{Representation of the \(K_{xx}\) component of the nodal field input.}
\label{tab:Kxx_nodal_data_analytic_field}
\begin{tabular}{ccccc}
\toprule
Node & \(x\) & \(y\) & \(z\) & Field value \\
\midrule
1 & 0      & 0      & 0 & 10.513 \\
2 & 1.1794 & 0      & 0 & 10.700 \\
3 & 0.0002 & 1.1185 & 0 & 10.663 \\
4 & 1.1797 & 1.1185 & 0 & 10.626 \\
\vdots & \vdots & \vdots & \vdots & \vdots \\ 
\bottomrule
\end{tabular}
\end{table}

The SCACS models considered here are obtained from Ref.~\cite{Ugwumadu2026SCACS}, and denoted \(M1\), \(M2\), and \(M3\), corresponding to a twin-grain-boundary Si nanowire, an amorphous--crystalline Si interface, and a Si nanopillar, respectively. The mesh files for the models were first converted to an unstructured-grid format (\texttt{.vtu}). Meshio \cite{Schlomer_meshio_Tools_for} was then used to convert the unstructured meshes into Abaqus input files (\texttt{.inp}).  The SCACS point-cloud conductivity fields were mapped to Abaqus following the procedure outlined in Ref. \cite{Abaqus_Mapping_fields}.

\begin{figure}[!t]
    \centering
    \includegraphics[width=\linewidth]{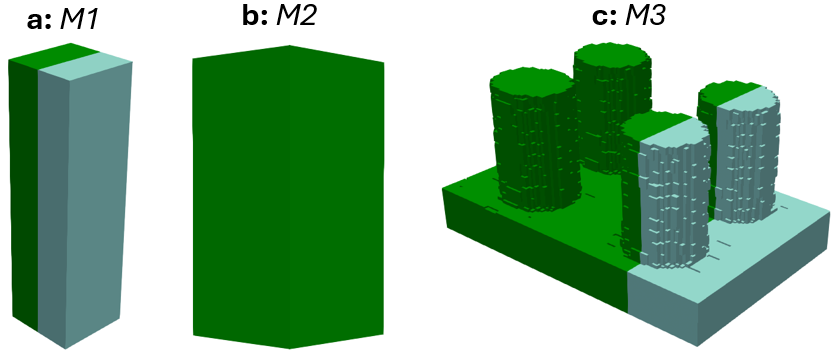}
    \caption{Geometrical representations of the models: (a) $M1$, (b) $M2$, and (c) $M3$. the green region show the slice considered in this work. The cyan region is omitted.}
    \label{Fig0_Slices}
\end{figure}

\begin{figure*}[ht!]
    \centering
    \includegraphics[width=\linewidth]{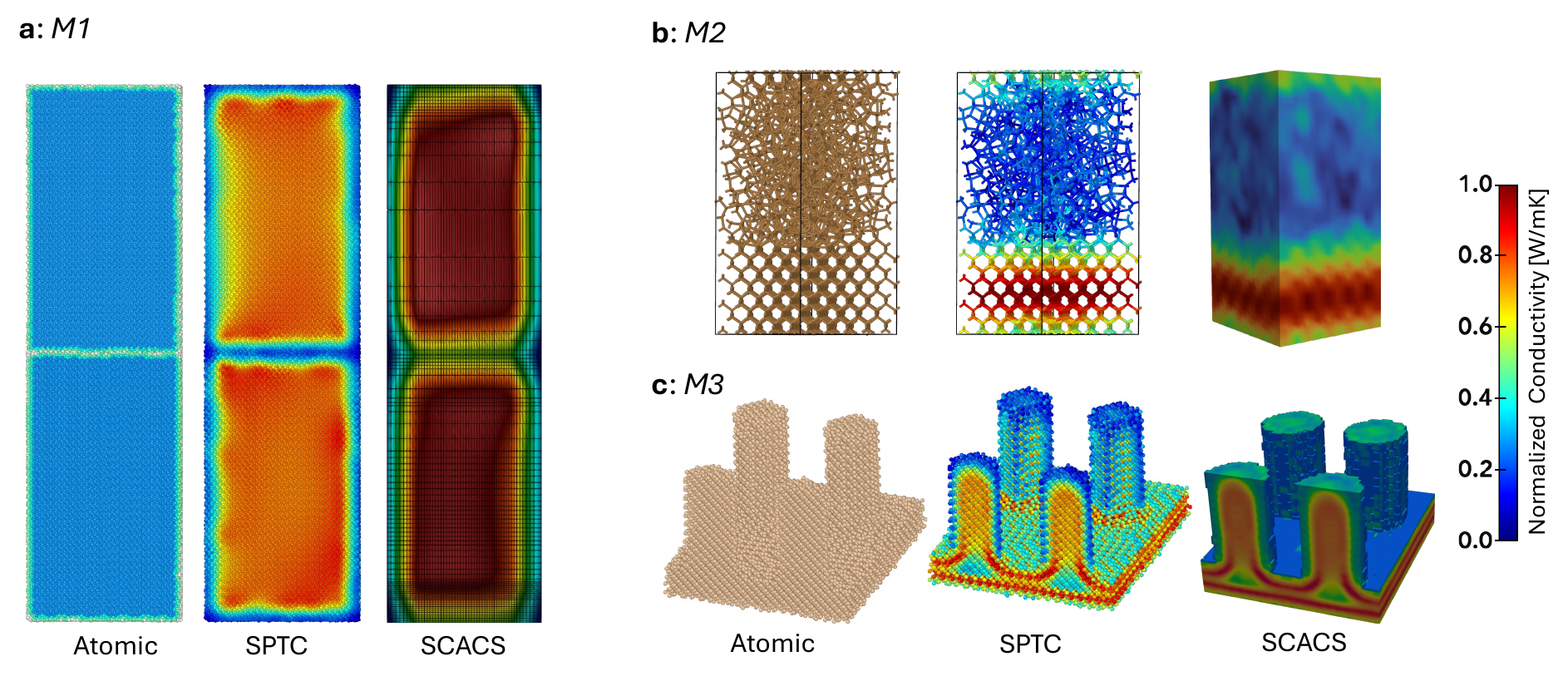}
    \caption{Multiscale representations of structures (a) $M1$, (b) $M2$, and (c) $M3$. Columns show, from left to right, the atomic representative volume element, the SPTC atom-resolved conductivity, and the SCACS coarse-grained finite-element conductivity field mapped to Abaqus.}
    \label{fig:Fig1_Models}
\end{figure*}

For the presentation and the discussion of the results in this work, the green region in Fig. \ref{Fig0_Slices}(a--c) highlights the slice and surface considered for each model (i.e., the cyan region is omitted).  Figure~\ref{fig:Fig1_Models}(a--c) illustrates the SCACS-to-Abaqus mapping for models $M1$, $M2$, and $M3$, respectively. A more detailed cross-section for $M3$ is shown in Fig. \textcolor{magenta}{S1} of the supplemental material. The columns show, from left to right, the atomic-scale RVE, SPTC atom-resolved thermal conductivity, and the SCACS coarse-grained finite-element conductivity field  mapped into Abaqus. All the continuum-scale models use DC3D8 hexahedral elements, with properties summarized in Table~\ref{tab:model_info}. Further discussion of the implementation of the mapping of the SCACS-based conductivity field Abaqus is provided in Appendix \ref{app:psuedoCode}. 

\begin{table}[h!]
\scriptsize
\renewcommand{\arraystretch}{1.4}
\centering
\caption{Mesh information for the models. The unit of the isotropic (scalar) effective conductivity $K^\mathrm{eff}$ is in $\mathrm{Wm^{-1}K^{-1}}$.}
\label{tab:model_info}
\begin{tabular}{lllll}
\toprule
Model & Elements & Nodes & $K^\mathrm{eff}$ & Description \\ 
\midrule
$M1$   & 1,682,384  & 162,000  & 33.20 & twin-grain-boundary Si nanowire  \\ 
$M2$   & 5,417,280 & 511,584 & 13.80 & amorphous–crystalline Si interface \\ 
$M3$   & 462,372 & 504,621 & 27.98 & Si nanopillar structure (4 pillars)\\ 
\bottomrule
\end{tabular}
\end{table}

After import, each mesh was translated in Abaqus \texttt{Assembly} module to align the mesh coordinates with the SCACS conductivity-field coordinates. For the heat-flow calculations, Dirichlet boundary conditions were applied by prescribing temperatures of 500 K and 300 K on the top and bottom faces, respectively. This matches the boundary-condition treatment used for SCACS native solver calculations in Ref.~\cite{Ugwumadu2026SCACS}. Results obtained under natural (Neumann) boundary conditions will be presented in a subsequent publication.

Finite-element simulations were performed using the steady-state and transient heat-transfer step type in Abaqus. Density was set to $\approx$ 1.2 kg/m$^3$  and the specific heat was set to 700 J/kg$\cdot$K (for the transient calculation). The finite-element formulation for the heat-flow problem is summarized in Appendix \ref{app:FE_formulation}. The direct solver and ``Full Newton" solution technique were used \cite{Abaqus2024Docs}. The requested field outputs were coordinates (COORD), predefined field values (FV), heat-flux vector (HFL), and nodal temperature (NT). These solutions were used to compare the SCACS-derived conductivity fields to uniform-scalar-conductivity assignments within Abaqus. 

The SCACS data-point files, the corresponding Abaqus input files, and the a \texttt{python} utility used for the conversion are provided as auxiliary data in Ref. \cite{Zenodo} and the description of the files are provided Tables \textcolor{magenta}{S1} (SCACS) and \textcolor{magenta}{S2} (Abaqus) in Section \textcolor{magenta}{S1} of the supplemental material \cite{SM}.

\begin{figure*}[!tbhp]
    \centering
    \includegraphics[width=.85\linewidth]{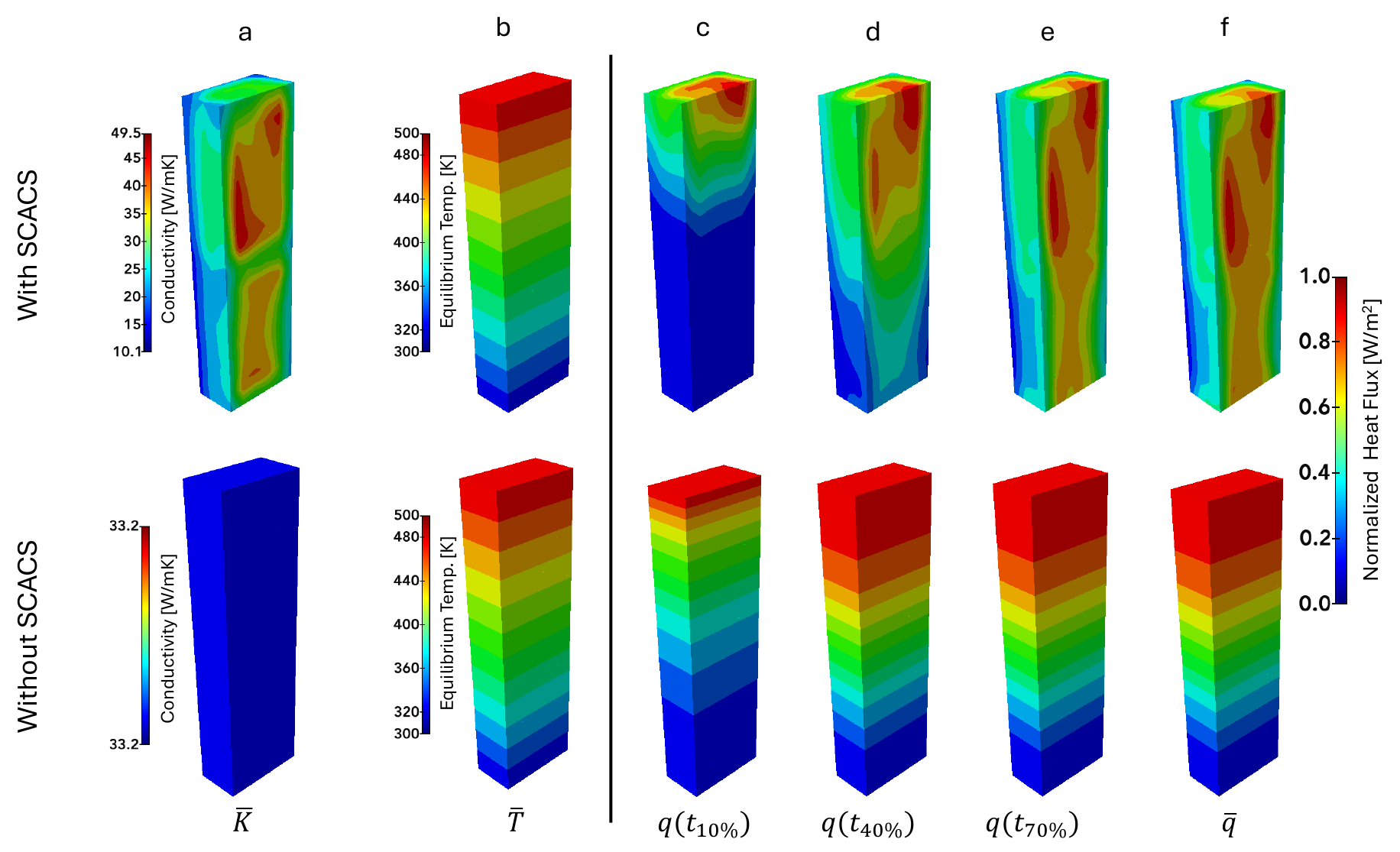}
    \caption{Analysis of the grain-boundary Si nanowire \(M1\). The colormaps show normalized values of (a) the isotropic conductivity field, $\overline K$, and (b) the steady-state temperature field, $\overline T$. Panels (c--e) show the transient heat-flux, $q_t$, at different times: (c) $t = 10\%$, (d) $t = 40\%$, and (e) $t = 70\%$ of the convergence time. The steady-state heat-flux, $\overline q$, is shown in (f).}
    \label{fig:Fig2_M1}
\end{figure*}

\section{Results}

We evaluate heat transport using SCACS-derived, Abaqus-mapped finite-element RVEs that preserve the spatially resolved conductivity information obtained from the underlying atomic structures. In homogenization, an RVE is a finite microscale domain whose averaged response represents the macroscopic material behavior. For periodic systems, the RVE is the repeating ``unit-cell"; for non-periodic systems, representativeness is established by convergence of the effective properties with increasing domain size~\cite{zienkiewicz2005finite,Geers2010}. In Ref.~\cite{Ugwumadu2026SCACS}, we showed that SCACS finite-element models constructed from atomistic data satisfy this RVE requirement.

Here, these RVEs are used to compute effective conductivity tensors, \(\overleftrightarrow{K}^{\mathrm{eff}}\), discussed in Appendix~\ref{app:Keff}. When a target or experimental conductivity \(k_t\) is available, the local elementwise conductivity tensors are uniformly rescaled [Eq.~\eqref{eq:K_scaled_element}], preserving the spatial variation and anisotropy encoded by the SCACS-derived field while matching the prescribed effective response. 

We compare the SCACS-derived anisotropic conductivity with conventional uniform-conductivity assignments in Abaqus. The comparison focuses on relative spatial and temporal variations in heat flow, rather than absolute steady-state differences between heterogeneous SCACS-derived conductivity fields and homogeneous single-scalar conductivity models. Accordingly, the results are shown as normalized fields for qualitative comparison, while quantitative effective conductivities are reported in Figures \textcolor{magenta}{S2}, \textcolor{magenta}{S3}, and \textcolor{magenta}{S4} for $M1$, $M2$, and $M3$, respectively in the Supplemental Material~\cite{SM}. 

The SCACS-derived isotropic effective conductivities, \(K^\mathrm{eff}\) [Eq.~\eqref{eq:K_eff_final_isotropic}], are listed in Table~\ref{tab:model_info} and used are as the uniform conductivity values in the conventional Abaqus models. The effective conductivities are approximately \(33.2\), \(13.8\), and \(28\) \(\mathrm{Wm^{-1}K^{-1}}\) for $M1$, $M2$, and $M3$, respectively. These values are subtantially lower than that of single-crystalline silicon at room temperature, \(125\pm19~\mathrm{Wm^{-1}K^{-1}}\), which decreases markedly to \(46\pm7~\mathrm{Wm^{-1}K^{-1}}\) at 425 $^\circ$C \cite{Morris1961_monocrystal_Si_cond}. Nevertheless, the conductivity range obtained for $M1$--$M3$ is consistent with the reduced thermal conductivities expected in nanostructured silicon containing grain boundaries, polycrystalline regions, amorphous phases, or surface- and geometry-induced structural reconstructions~\cite{Isotta2024_GB_Si,BenAmor2019,Takeuchi2020PolySi,Juangsa_SiNP_Cond,DavisNanopillars2014}.

\subsection{Heat-flow response of the grain-boundary Si nanowire}

Figures~\ref{fig:Fig2_M1} and \textcolor{magenta}{S2} compare the thermal response of the grain-boundary Si nanowire, $M1$, using SCACS-derived and uniform conductivity fields, shown in the upper and lower rows, respectively. The SCACS-derived field in Fig.~\ref{fig:Fig2_M1}(a) exhibits pronounced spatial heterogeneity, with local conductivity values ranging from approximately \(10.1\) to \(49.5~\mathrm{Wm^{-1}K^{-1}}\). Extended regions of relatively high conductivity occur primarily through the interior and along one side of the nanowire, while lower-conductivity regions appear near portions of the opposing surface and around structurally perturbed regions. These variations reflect the influence of the grain boundary and local atomic disorder on thermal transport [Fig. \ref{fig:Fig1_Models}(a)]. By contrast, the uniform-conductivity model assigns a single effective conductivity value throughout the domain and therefore contains no information about local transport pathways.

\begin{figure*}[!tbhp]
    \centering
    \includegraphics[width=\linewidth]{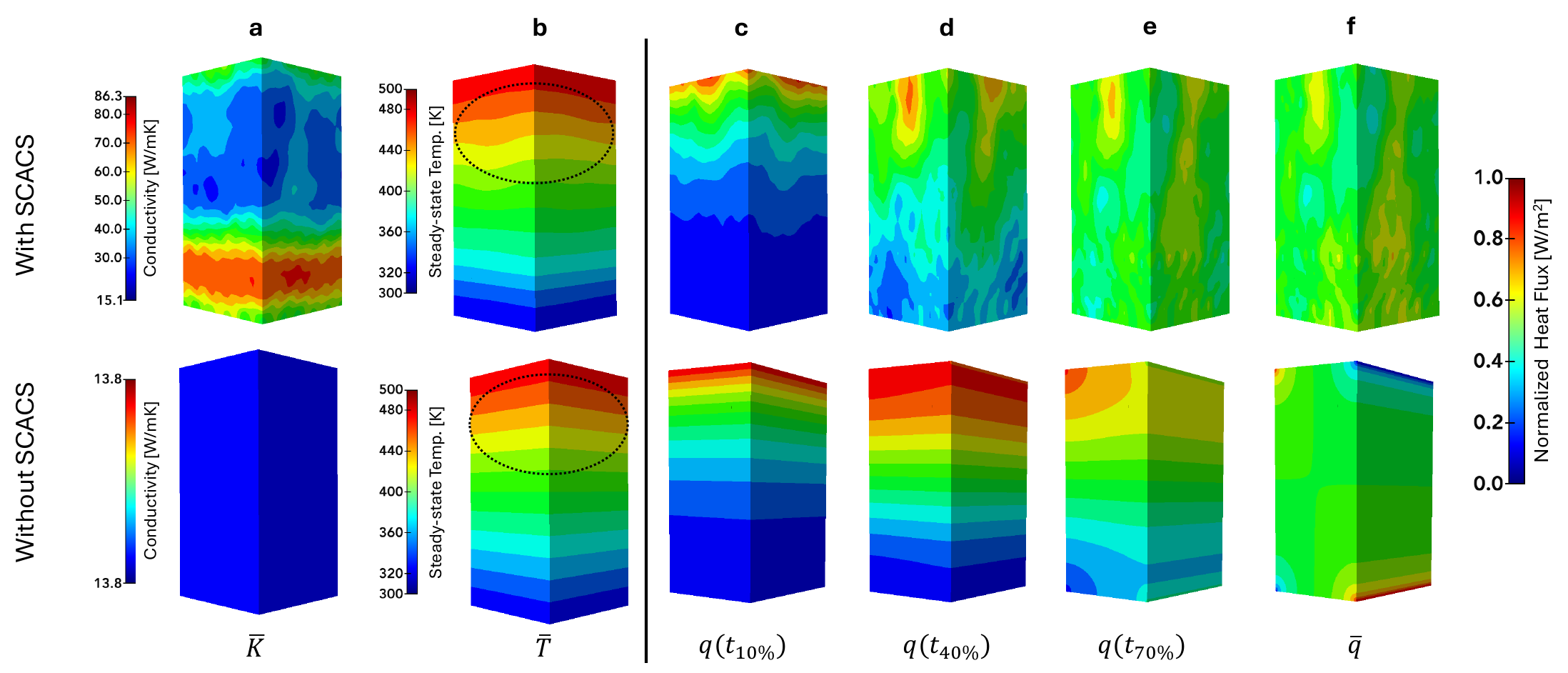}
    \caption{Analysis of the amorphous-crystalline Si interface structure (M2). The colormaps show normalized values of (a) the isotropic conductivity field, $\overline K$, and (b) the steady-state temperature field, $\overline T$. Panels (c--e) show the transient heat-flux, $q_t$, at different times: (c) $t = 10\%$, (d) $t = 40\%$, and (e) $t = 70\%$ of the convergence time. The steady-state heat-flux, $\overline q$, is shown in (f).}
    \label{fig:Fig3_M2}
\end{figure*}

The effect of this heterogeneity is not strongly apparent in the steady-state temperature field in Fig.~\ref{fig:Fig2_M1}(b). In both the SCACS-based and uniform models, the temperature varies primarily along the axial direction, producing nearly planar isothermal layers. As shown later for \(M2\) and \(M3\), however, this behavior is not universal. Despite having the same homogenized conductivity, the two models yield different internal transport patterns, as revealed by the heat-flux fields in Fig.~\ref{fig:Fig2_M1}(c--f). 

The transient heat-flux fields in Fig.~\ref{fig:Fig2_M1}(c--e) show how the atom-informed conductivity controls the development of heat-flow pathways. At \(t=10\%\) of the convergence time, the thermal disturbance remains concentrated near the heated boundary. In the SCACS model, however, the advancing flux front is nonuniform and penetrates more rapidly through locally conductive regions. As the system evolves to \(t=40\%\), distinct high-flux channels emerge and extend farther along the nanowire. By \(t=70\%\), these pathways span most of the domain and approach the heterogeneous pattern observed at steady state. The transient response is therefore not simply an axially translating planar front; it is continually redirected by the underlying conductivity landscape.

In contrast, the uniform-conductivity model produces horizontally layered flux fields at all three time snapshots. The flux front advances through the nanowire with little lateral variation because every element presents the same thermal resistance. This behavior is mathematically consistent with the homogeneous material assignment but cannot represent the preferential pathways and local bottlenecks associated with the grain-boundary microstructure. The difference between the two transient solutions also indicates that matching only \(K^{\mathrm{eff}}\) does not ensure an equivalent thermal timescale or local energy-transport response.

At steady state [Fig.~\ref{fig:Fig2_M1}(f)], the uniform model approaches an almost spatially constant normalized heat flux, whereas the SCACS-based model retains strongly nonuniform transport. The largest flux follows connected high-conductivity regions, while reduced flux occurs where the local conductivity is smaller or where the heat-flow path is redirected. This persistent spatial correlation between \(\overline K\) and \(\overline q\) is consistent with Fourier's law [Eq.~\eqref{eq:Fourier}], but it also demonstrates that the heat-flux pattern cannot be inferred from conductivity alone: the temperature gradient adapts to the heterogeneous field through the global finite-element solution.

\subsection{Heat-flow response of the amorphous--crystalline Si interface}

Figures~\ref{fig:Fig3_M2} and \textcolor{magenta}{S3} presents the heat-transport behavior of the amorphous--crystalline Si interface structure, $M2$, for the SCACS-based and conventional uniform-conductivity implementation in the upper and lower rows, respectively. The conductivity field in Fig.~\ref{fig:Fig3_M2}(a) is strongly heterogeneous, varying from approximately \(1.5\) to \(8.6~\mathrm{Wm^{-1}K^{-1}}\). The field contains extended high-conductivity regions near the lower (crystalline) portion of the structure and spatially irregular lower-conductivity domains in the upper and lateral (amorphous) regions  [Fig. \ref{fig:Fig1_Models}(b)]. In contrast, the homogeneous conductivity model assigns the single effective conductivity value  (Table \ref{tab:model_info}) throughout the domain and therefore cannot distinguish the amorphous, interfacial, and crystalline regions.

\begin{figure*}[!tbhp]
    \centering
    \includegraphics[width=\linewidth]{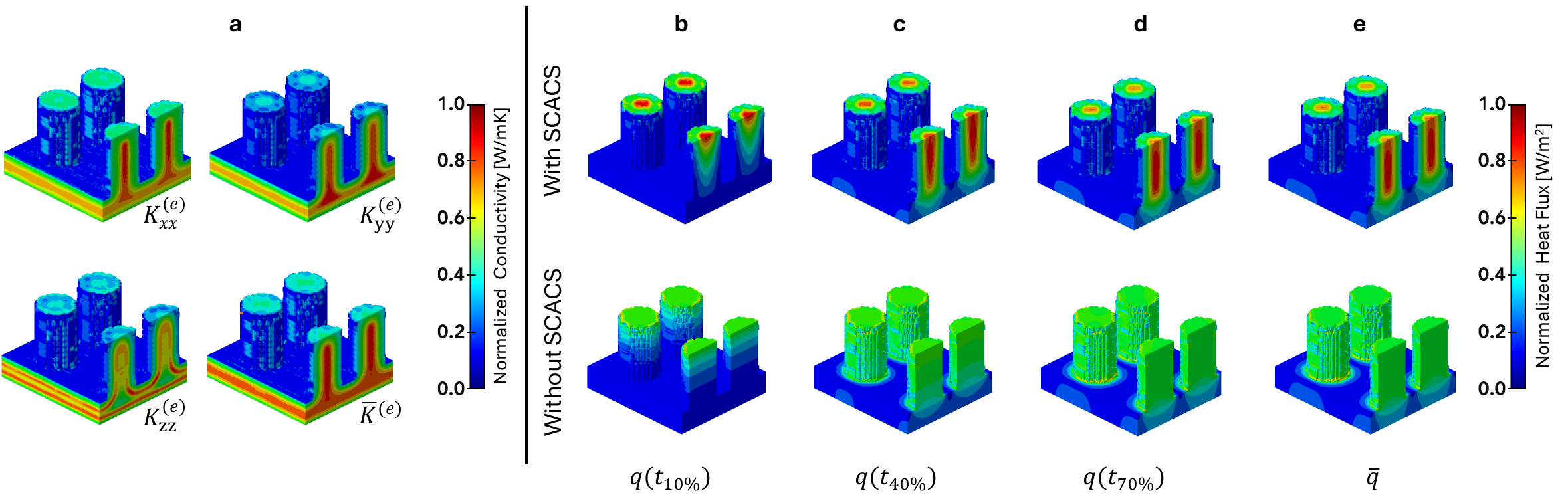}
    \caption{Analysis of the Si nanopillar structure (\(M3\)). Panel (a) shows the normalized elementwise directional conductivity fields, \(K_{xx}^{(e)}\), \(K_{yy}^{(e)}\), and \(K_{zz}^{(e)}\), together with the corresponding isotropic conductivity field, \(\overline K^{(e)}\). Panels (b--d) show the normalized transient heat-flux field, \(q_t\), at (b) \(t=10\%\), (c) \(t=40\%\), and (d) \(t=70\%\) of the convergence time. Panel (e) shows the normalized steady-state heat-flux field, \(\overline q\). The upper and lower rows correspond to simulations performed with and without the SCACS-derived conductivity field, respectively.}
    \label{fig:Fig4_M3}
\end{figure*}

The consequence of this material heterogeneity is visible in the steady-state temperature fields in Fig.~\ref{fig:Fig3_M2}(b). The uniform model produces nearly planar isotherms and an approximately one-dimensional temperature gradient between the prescribed hot and cold boundaries. The SCACS-based temperature field is markedly different. Its isotherms are curved and spatially redistributed, particularly in the upper part of the model, as highlighted by the dashed region. This departure from a layered temperature profile indicates that the local temperature gradient adapts to the spatial conductivity landscape. Regions with lower conductivity sustain larger temperature gradients, whereas more conductive regions permit heat to pass with a smaller local temperature drop.

This behavior is physically intuitive as interfaces are not generally geometrically sharp and uniform thermal boundary, but instead, are finite regions with spatially varying transport properties inherited from the underlying atomic arrangement. The resulting temperature field therefore captures local thermal resistance and lateral redistribution that would be removed by assigning a single effective conductivity to the full model. Although the homogeneous model may reproduce an averaged end-to-end thermal response, it does not preserve the local temperature variations that are relevant to hot-spots, thermal stress, and degradation.

The transient heat-flux fields in Fig.~\ref{fig:Fig3_M2}(c--e) provide further evidence of this distinction. At \(t=10\%\) of the convergence time, the heat flux in both models is concentrated near the hot boundary. In the uniform model, the advancing thermal front remains approximately planar and follows the imposed axial temperature gradient. In the SCACS model, the front is visibly distorted and progresses at different rates across the cross section. The flux penetrates more rapidly through connected higher-conductivity regions, while low-conductivity zones delay the downward propagation of heat.

At \(t=40\%\), the SCACS-derived heat-flux develops a complex network of preferential transport pathways. Regions of enhanced flux extend through portions of the structure, while neighboring low-flux areas persist, particularly near the lower and interfacial regions. By \(t=70\%\), the overall flux magnitude becomes more evenly distributed, but the spatial imprint of the heterogeneous conductivity field remains visible. The transient response therefore evolves through microstructure-dependent pathways rather than through the uniformly layered progression observed in the uniform-conductivity model.

The contrast becomes particularly clear in the steady-state heat-flux field shown in Fig.~\ref{fig:Fig3_M2}(f). The SCACS-informed model retains substantial spatial variation, with higher flux following locally conductive regions and reduced flux occurring in areas of greater thermal resistance. The uniform model instead predicts broad, smooth regions of nearly constant flux, with variations occurring mainly near boundaries and geometric corners. These localized features are primarily consequences of the imposed boundary conditions and finite geometry (edge effects) rather than of any internal material structure.

\subsection{Heat-flow response of the Si nanopillar structure}

Figures~\ref{fig:Fig4_M3} and \textcolor{magenta}{S4} show the Si nanopillar structure, \(M3\), modeled with SCACS-derived conductivity and the uniform reference field. The normalized elementwise components \(K_{xx}^{(e)}\), \(K_{yy}^{(e)}\), \(K_{zz}^{(e)}\), and \(\overline K^{(e)}\) are shown in Fig.~\ref{fig:Fig4_M3}(a). Variations among the directional components indicate local anisotropy, whereas the isotropic field captures only the overall spatial pattern. Conductivity is higher within selected regions of the pillars and base and lower near several surfaces and reconstructed areas, showing that a single conductivity value cannot adequately describe either region.

The upper and lower rows in Fig.~\ref{fig:Fig4_M3}(b--e) correspond to the SCACS-based and homogeneous conductivity models, respectively. The influence of this heterogeneous conductivity field is evident in the transient heat-flux distributions. At \(t=10\%\) of the convergence time [Fig.~\ref{fig:Fig4_M3}(b)], heat flow is concentrated near the thermally driven surfaces. In the SCACS model, the flux is highly localized, with the largest values appearing near the tops of selected pillars and along conductive regions close to their sidewalls. The four pillars do not exhibit identical flux distributions, even though they belong to the same nominal structure. This asymmetry reflects differences in their local conductivity fields as the thermal loading was uniformly imposed.

In contrast, the uniform-conductivity model produces similar flux magnitudes and distributions across pillars having comparable geometry. The heat flux is distributed more smoothly around the pillar surfaces and through the base because every finite element is assigned the same thermal resistance. Consequently, the homogeneous model responds primarily to geometric variations, whereas the SCACS-informed model responds to both geometry and the atomic-level material heterogeneity.

As the transient solution evolves to \(t=40\%\) and \(t=70\%\) [Fig.~\ref{fig:Fig4_M3}(c,d)], heat penetrates farther into the pillars and the supporting base. In the SCACS calculation, the initially localized regions develop into persistent preferential pathways. High-flux channels remain concentrated within specific portions of the pillars, particularly in regions that coincide with enhanced directional conductivity. Neighboring regions remain weakly conducting and carry considerably less heat. The spatial pattern therefore becomes established early in the transient response and is retained as the system approaches equilibrium. 

The uniform-conductivity model evolves differently. Although its flux increases throughout the pillars and spreads into the base, the distributions remain comparatively smooth and similar among equivalent geometric features. Local enhancements appear mainly near pillar--base junctions, edges, and curved surfaces, where the geometry redirects the temperature gradient. 

At steady state [Fig.~\ref{fig:Fig4_M3}(e)], the SCACS model retains pronounced spatial variations in heat flux. The largest normalized flux occurs within connected regions extending through selected pillars, while other parts of the same pillars carry substantially less heat. This result demonstrates that the pillars do not operate as uniformly conducting channels. Instead, their effective contribution to heat transport is controlled by internal pathways inherited from the local atomic structure. The supporting base similarly contains nonuniform transport regions that redistribute heat between the pillars and the external boundaries.

The uniform-conductivity result lacks this internal selectivity. Its steady-state flux is governed predominantly by the dimensions, orientations, and connectivity of the pillars. While such a model may reproduce the overall effective thermal response, it predicts nearly equivalent transport through geometrically similar features and cannot identify which regions preferentially carry heat or act as local bottlenecks.

\section{Discussion}

The results for $M1$, $M2$, and $M3$ establish three broader implications of implementing SCACS-derived conductivity fields in heat-flow problems. First, atomistic transport information can be incorporated into any continuum solver (solver-agnostic). Second, the spatially heterogeneous conductivity produces thermal behavior that cannot be recovered from a single effective property. Third, direct field mapping provides a physically consistent basis for simulations of refined and geometrically complex structures. We address each of these points in the subsections below.

\subsection{Bridging Atomistic Information and Continuum Workflows}

A central result of this work is that SCACS-derived conductivity fields are not restricted to the native SCACS finite-element solver introduced in the foundational study in Ref. \cite{Ugwumadu2026SCACS}. The conductivity information was transferred into Abaqus as spatially varying directional fields and used within its standard heat-transfer framework. SCACS provides the conductivity field, while Abaqus performs the finite-element assembly, solution, and post-processing.

Separating atom-informed property generation from the continuum solver allows SCACS fields to be integrated into established engineering workflows without changing the user’s preferred platform. Using native Abaqus capabilities for spatially varying orthotropic properties, the method was applied successfully to all three models and preserved their atomistically-derived heterogeneity in both transient and steady-state solutions. This establishes SCACS as a solver-agnostic material-field framework that can be used in any finite-element platform supporting spatially varying conductivity tensors and material orientations.

It is worth noting that the present implementation assumes that the principal conductivity directions remain aligned with the global finite-element axes and that off-diagonal conductivity components are negligible. More general systems may require local material rotations or fully populated conductivity tensors. Nevertheless, the current results establish the practical feasibility of transferring atom-derived anisotropic material information into any continuum solver.

\subsection{Improved Heat-Flow Prediction from Spatially-resolved Conductivity Fields}

The comparison with the uniform-conductivity models confirms that matching the homogenized conductivity does not guarantee the same internal thermal response. In all three structures, the SCACS-based and uniform-conductivity models used the same geometry, boundary conditions, and effective conductivity. The differences in their solutions therefore arise from the spatial and directional variations retained in the SCACS fields.

For \(M1\), the steady-state temperature distributions remain broadly similar because both models are dominated by the imposed axial temperature difference. The heat-flux fields, however, differ substantially. The SCACS model develops localized transport channels and low-flux regions associated with the grain boundary, surfaces, and local structural disorder, whereas the uniform model produces a smoother and more layered response. This indicates that a similar macroscopic temperature field can conceal important differences in the internal transport pathways.

The effect of heterogeneity is more apparent in \(M2\). The amorphous, interfacial, and crystalline regions are represented by a continuous spatial distribution of conductivity rather than by a single uniform property. This produces curved isotherms, distributed thermal resistance, and nonuniform transient heat-flow pathways. The uniform model cannot distinguish these structural regions and therefore predicts a smoother, nearly one-dimensional response.

The \(M3\) nanopillar results further separate the effects of geometry from those of material heterogeneity. Both models capture geometric features such as pillar edges, surfaces, and pillar--base junctions. However, the uniform model treats geometrically similar regions as thermally equivalent. The SCACS-derived conductivity model reveals internal variations within each pillar, asymmetric heat-flow behavior among the pillars, and persistent transport bottlenecks. Differences among the directional conductivity fields also show that the local response cannot always be represented by a single isotropic value.

These results demonstrate that effective conductivity alone is insufficient when local thermal behavior is important. A uniform property may reproduce the overall boundary-to-boundary response, but it suppresses local gradients, preferential pathways, and thermally resistive regions. Such information is relevant to hot-spot formation, thermal stress, interfacial degradation, and temperature-sensitive device performance.

The results also show that the influence of heterogeneity depends on the structure. In some systems, such as \(M1\), it is more visible in the heat-flux field than in the temperature field. In others, such as \(M2\), it modifies both. The importance of spatially resolved conductivity therefore depends on the microstructure, geometry, loading direction, and contrast within the conductivity field.

\subsection{Physically Consistent Field Mapping in Refined Meshes}

The SCACS-to-Abaqus procedure provides a systematic alternative to manually assigning material properties after mesh generation. In a conventional workflow, refining a mesh does not add new physical information when the same uniform conductivity is assigned to every element \cite{zienkiewicz2005finite,Ugwumadu2025SPTC}. It only represents the same homogeneous material using a larger number of elements.

In the SCACS approach, the conductivity field is defined spatially from the underlying atomistic model and mapped onto the continuum mesh. Each region of the finite-element domain therefore retains conductivity information associated with its local atomic environment. Grain boundaries, interfaces, surfaces, and the atomic conformations enter the continuum model through the material field rather than through manually selected partitions or fitted property values.

This consideration may bear importance for some structures than for others, for example, \(M2\) and \(M3\) exhibit conductivity variations that do not coincide with simple geometric boundaries. The amorphous--crystalline interface is represented as a finite transition region rather than an idealized sharp boundary. Similarly, the nanopillars contain internal conductivity variations that cannot be captured by assigning one property to the pillars and another to the base.

The coarse-graining procedure also produces a smoothly varying field. This avoids abrupt element-to-element changes that may result from direct nearest-neighbor assignment or manual phase labeling. A smooth field better represents the gradual structural variations present in the atomic model and reduces the likelihood of numerical features caused solely by artificial material discontinuities.

\section{Conclusion and Outlook}

We demonstrated that SCACS-derived thermal conductivity fields can be transferred into Abaqus and used within a standard finite-element heat-transfer workflow. Across the grain-boundary nanowire, amorphous--crystalline interface, and nanopillar structures, the mapped fields preserved the spatial heterogeneity and directional dependence inherited from the underlying atomistic models.

The SCACS-informed models revealed localized heat-flow pathways, distributed interfacial resistance, anisotropic transport, and geometry-dependent bottlenecks that were absent from the homogeneous models. These differences were most evident in the transient and steady-state heat-flux fields and, for the amorphous--crystalline interface, in the temperature field. The results establish SCACS as a solver-agnostic material-field framework for introducing atom-informed conductivity into existing engineering simulations.

Future work will extend the present approach to fully populated (off-diagonal) conductivity tensors, spatially varying material orientations, and coupled thermal, electrical, and mechanical problems. Quantitative mesh-convergence and computational-cost studies are also needed to determine whether conductivity- or residual-based adaptive refinement can reduce the number of elements required for a given accuracy. Additional validation against experiments and other continuum solvers will further establish the transferability and predictive value of the framework. Ultimately, this approach provides a practical route for incorporating atomistic materials information into engineering-scale simulations for device design, extreme-environment applications, and multiscale materials prediction.

\begin{acknowledgments}
This work is funded by the Laboratory Directed Research and Development (LDRD) program at Los Alamos National Laboratory through the Director’s Postdoctoral Fellowship Program (Project No. 20240877PRD4) and the Exploratory Research (ER) program (Project No. 20240397ER). Los Alamos National Laboratory is operated by Triad National Security, LLC, for the U.S. Department of Energy (DOE) National Nuclear Security Administration under Contract No. 89233218CNA000001. 
\end{acknowledgments}

\section*{Data Availability}
The data that support the findings of this article are openly available \cite{Zenodo}. The source code for \texttt{SCACS} will be available upon request subject to LANL and US export control policies. 

\section*{Author Contributions}
C.U. and R.M.T conceived the idea. W. D. and C.U. implemented the idea and prepared the manuscript. M.A and R.M.T revised the manuscript.

\begin{appendix}

\section{Protocol for mapping SCACS-based conductivity field to Abaqus}\label{app:psuedoCode}

The spatially varying orthotropic conductivity was implemented using Abaqus predefined and analytic field variables. The atomistically derived conductivity field was defined by three principal conductivity components, $K_{xx},K_{yy},$ and $K_{zz}$, evaluated throughout the finite element geometry. These values were mapped into Abaqus as three independent field variables, $
F_1(\mathbf{r}) \rightarrow K_{xx}(\mathbf{r})$, $F_2(\mathbf{r}) \rightarrow K_{yy}(\mathbf{r})$, and
$F_3(\mathbf{r}) \rightarrow K_{zz}(\mathbf{r})$,
where $\mathbf{r}$ denotes the spatial position in the finite element domain. The field variables provide the geometric and spatial mapping of the conductivity values, while the field-dependent orthotropic conductivity definition maps those values to the diagonal terms of the conductivity tensor.

At each integration point, Abaqus evaluates the local field-variable state,
\(
\mathbf{F}(\mathbf{r}) =
\left\{
F_1(\mathbf{r}),
F_2(\mathbf{r}),
F_3(\mathbf{r})
\right\},
\)
and uses a field-dependent material definition to assign the conductivity components, $K(F_1, F_2, F_3)$.

The local conductivity tensor is then assigned as
\begin{equation}
\mathbf{K}_{\mathrm{local}} =
\begin{bmatrix}
K_{xx} & 0 & 0 \\
0 & K_{yy} & 0 \\
0 & 0 & K_{zz}
\end{bmatrix},
\end{equation}

The local orthotropic axes were aligned with the global finite element coordinate system using an identity orientation matrix, $\mathbf{R}$, so that the global conductivity tensor can be obtained as
\begin{equation}\label{eq:K_global}
\mathbf{K}_{\mathrm{global}} = \mathbf{R}
\mathbf{K}_{\mathrm{local}}
\mathbf{R}^{T},
\end{equation}
hence, $K_{xx},K_{yy},$ and $K_{zz}$ correspond directly to the global $X, Y,$ and $Z$ directions, respectively. The implementation in Abaqus is summarized as follows:

\begin{algorithm}[H]
\small
\caption{Workflow for mapping SCACS-derived conductivity fields in Abaqus}
\label{alg:orthotropic_conductivity_workflow}
\begin{algorithmic}[1]
\State Import SCACS point-cloud data with spatial coordinates $\mathbf{r}=x_i,y_i,z_i$, and principal conductivity components $\mathbf{K}(\mathbf{r})=[K_{xx}(\mathbf{r}),K_{yy}(\mathbf{r}),K_{zz}(\mathbf{r})]$.
\State Extract the spatial conductivity $K_{\alpha \alpha}(\mathbf{r})$; $\alpha \in \{x,y,z\}$ into Abaqus field variables: $F_1(\mathbf{r})= K_{xx}(\mathbf{r}), F_2(\mathbf{r})=K_{yy}(\mathbf{r}),$ and   $F_3(\mathbf{r})=K_{zz}(\mathbf{r})$.
\State Define the field-dependent orthotropic conductivity table such that $[K_{xx},K_{yy},K_{zz}]=K(F_1,F_2,F_3)$.
\State Assign the orthotropic material definition to the finite element domain.
\State Preserve the local-to-global conductivity mapping using Eq. \eqref{eq:K_global}.
\State Apply the boundary conditions and thermal loading, and evaluate the heat-flux and temperature.
\end{algorithmic}
\end{algorithm}

\section{The Finite Element Equation}\label{app:FE_formulation}

From the continuity and constitutive relations for the heat-flow problem in Equations \eqref{eq:continuityRelation} and \eqref{eq:Fourier}, respectively, we define the computational domain $\Omega\subset\mathbb R^3$, and the boundary
\(
\partial\Omega = \Gamma_D \,\dot\cup\, \Gamma_N \,\dot\cup\, \Gamma_R ,
\) where $\Gamma_D$, $\Gamma_N$, and $\Gamma_R$ correspond to Dirichlet, Neumann, and Robin boundaries, respectively. The prescribed boundary conditions with a volumetric source $f(\mathbf x)$ are
\begin{align}
 \nabla^\mathsf{T} \big(\overleftrightarrow{K}(\mathbf x)\nabla T \big) &= f(\mathbf x)  && \text{in } \Omega, \label{eq:strongForm}\\
 T &= \widetilde T && \text{on } \Gamma_D,\\
 q_n  &= \widetilde q && \text{on } \Gamma_N,\\
  h\,(T - T_\infty) &= -q_n && \text{on } \Gamma_R,
\end{align}
where $\widetilde T$ is the prescribed temperature, $h$ is the heat transfer or radiation coefficient, $T_\infty$ is the ambient temperature, $\widetilde q$ is a specified heat-flux value, and
\begin{equation}\label{eq:normal_flux}
    q_n = \mathbf n^\mathsf{T} \big(\overleftrightarrow{K}(\mathbf x)\nabla T\big),
\end{equation} 
with $\mathbf n$ the outward unit normal on the surface boundary.

The weak formulation of Eq. \eqref{eq:strongForm} is given as \cite{zienkiewicz2005finite}
\begin{equation}
\label{eq:Continous_WeakForm}
\begin{aligned}
&\int_{\Omega} 
\nabla v^\mathsf{T}
\bigl(\overleftrightarrow{K}(\mathbf{x}) \nabla T\bigr)
\,\mathrm{d}\Omega
+
\int_{\Gamma_R} h\,v\,T\,\mathrm{d}\Gamma
\\
&\; =
-\int_{\Omega} v\,f(\mathbf{x})\,\mathrm{d}\Omega
+
\int_{\Gamma_N} v\,\widetilde{q}\,\mathrm{d}\Gamma
+
\int_{\Gamma_R} h\,T_{\infty}\,v\,\mathrm{d}\Gamma .
\end{aligned}
\end{equation}
valid for all test functions $v \in \mathcal{V} := \{v \in H^1(\Omega)\,|\, v = 0 \text{ on }\Gamma_D\}$, where $H^1(\Omega)$ is the Sobolev space of square-integrable functions with square-integrable gradients \cite{Adams2003sobolev}.  The finite element solution follows the standard Galerkin approach \cite{zienkiewicz2005finite} where on each element $e$, we approximate the temperature field as
\begin{equation}
T \approx T^h = \sum_{b\in\mathcal I_e} N_b\,T_b,
\end{equation}
where $N_b$ are $C^0$ shape functions \cite{zienkiewicz2005finite}, $T_b$ are the nodal temperatures and $\mathcal I_e$ is the set of node indices. The corresponding test function on $e$ is taken as $v = N_a$, so that the elementwise heat-flow equation becomes
\begin{equation}
\begin{aligned}\label{eq:FE_WeakForm}
&\sum_{b\in\mathcal I_e}
\left[
  \int_{\Omega^e} \overrightarrow{G}_a^\mathsf{T} \,\overleftrightarrow{K}^{(e)}\,\overrightarrow{G}_b\,\mathrm{d}\Omega
  \;+\;
  \int_{\Gamma_R^e} h\,N_a\,N_b\,\mathrm{d}\Gamma
\right] T_b
\\
&\; =
-\int_{\Omega^e} N_a\,f\,\mathrm{d}\Omega
\;+\;
\int_{\Gamma_N^e} N_a\,\widetilde{q}\,\mathrm{d}\Gamma
\;+\;
\int_{\Gamma_R^e} h\,T_{\infty}\,N_a\,\mathrm{d}\Gamma,
\end{aligned}
\end{equation}
for all $a\in\mathcal I_e$, where $\overrightarrow{G}_a = \nabla N_a$, and $\overleftrightarrow{K}^{(e)}$ is the SCACS-defined anisotropic conductivity tensor on $e$ in the element domain  \(\Omega_e\). Thus, each element
has its own conductivity tensor, and its directional components are
\(K_{xx}^{(e)}\), \(K_{yy}^{(e)}\), and \(K_{zz}^{(e)}\). A description of these quantities and their dimensions are provided in Table \textcolor{magenta}{S3} of the Supplemental Material \cite{SM}.

\section{Homogenization and Obtaining the Effective Conductivity}\label{app:Keff}
For an RVE of length \(D_\alpha\) in the loading direction \(\alpha \in \{x,y,z\}\), we solves a steady-state
two-face Dirichlet problem. Temperatures \(T_L\) and \(T_R\) are
prescribed on the left and right (opposing) boundary faces normal to \(\hat{\boldsymbol{\alpha}}\),
and no internal heat source is applied, \(f(\mathbf{x})=0\). The applied
temperature difference is
\begin{equation}
\Delta T = T_R - T_L.
\end{equation}
The effective conductivity along \(\alpha\) is first computed
from the boundary fluxes as
\begin{equation}
\label{eq:K_eff_flux}
K^\mathrm{eff}_{\alpha \alpha}
=
\frac{\overline{|Q|}}{\overline{A}}
\frac{D_\alpha}{|\Delta T|},
\end{equation}
where
\begin{equation}
\overline{|Q|}
=
\frac{|Q_L|+|Q_R|}{2},
\qquad
\overline{A}
=
\frac{A_L+A_R}{2}.
\end{equation}
Here, \(Q_L\) and \(Q_R\) are the total normal heat-fluxes through the
left and right driven boundaries, and \(A_L\) and \(A_R\) are the
corresponding boundary-face areas. For a driven boundary
\(\gamma\in\{L,R\}\),
\begin{equation}
Q_\gamma
=
\int_{\gamma} q_n\,\mathrm{d}\gamma,
\end{equation}
where \(q_n\) is defined in Eq. \eqref{eq:normal_flux}.

Using these elementwise tensors, the volume-averaged effective
conductivity in direction \(\alpha\) is computed as
\begin{equation}
\label{eq:K_eff_vol_sum}
K^\mathrm{vol}_{\alpha}
=
\frac{D_\alpha}{\Delta T\,|\Omega|}
\sum_{e=1}^{n_e}
\int_{\Omega_e}
\left(
\overleftrightarrow{K}^{(e)}
\nabla T
\right)
\cdot
\hat{\boldsymbol{\alpha}}
\,\mathrm{d}\Omega .
\end{equation}
where \(\Omega\) is the full RVE domain, \(|\Omega|\) is its volume,
\(\Omega_e\) is the domain of element \(e\), \(n_e\) is the total number of finite elements, and
\(\hat{\boldsymbol{\alpha}}\) is the unit vector in the loading direction.
The quantity \(K^\mathrm{vol}_{\alpha}\) is a single scalar effective
conductivity for the full RVE in direction \(\alpha\), not an elementwise
quantity.

If a target, or experimental, bulk conductivity \(k_\mathrm{t}\) is
specified, each element conductivity tensor is uniformly rescaled as
\begin{equation}\label{eq:K_scaled_element}
\overleftrightarrow{K}^{(e)}_\mathrm{s,\alpha}
=
\frac{k_\mathrm{t}}
{K^\mathrm{vol}_{\alpha}}
\overleftrightarrow{K}^{(e)},
\end{equation}
where \(e=1,\ldots,n_e\). Subsequent finite-element solves then use the scaled tensor \(\overleftrightarrow{K}^{(e)}_\mathrm{s,\alpha}\) for direction $\alpha$. Because the heat equation is linear
in the conductivity tensor, this scaling changes the magnitude of the
effective response while preserving the relative spatial variation and
anisotropy of the original SCACS-derived conductivity field.

The final homogenized conductivity in the loading direction is therefore
the single RVE-level value
\begin{equation}
\label{eq:K_eff_final}
K^\mathrm{eff}_{\alpha \alpha} = \overleftrightarrow{K}^{(e)}_\mathrm{s,\alpha}
\approx
k_\mathrm{t},
\end{equation}
up to discretization error. Repeating the procedure for
\(\alpha=x,y,z\) gives the directional effective conductivities:
\(
K^\mathrm{eff}_{xx},\;
K^\mathrm{eff}_{yy},\;
K^\mathrm{eff}_{zz}\), thus, giving a diagonal effective tensor, \(\overleftrightarrow{K}^{\mathrm{eff}}\), and the isotropic (scalar) effective conductivity is then 
\begin{equation}
\label{eq:K_eff_final_isotropic}
K^\mathrm{eff} = \frac{K^\mathrm{eff}_{xx} + K^\mathrm{eff}_{yy}+ K^\mathrm{eff}_{zz}}{3}.
\end{equation}

\end{appendix}

\bibliography{references}

\end{document}


\title{\Large {\bf Supplemental Material for}\\
`Solver-Agnostic Implementation of Atom-Informed Thermal Conductivity Fields in Continuum Heat-Flow Simulations'}

\author[1]{W. Downs}
\author[2]{C. Ugwumadu\thanks{cugwumadu@lanl.gov}}
\author[1]{M. Ali}
\author[2]{R. M. Tutchton}

\affil[1]{Department of Mechanical Engineering, Center for Advanced Materials Processing, Ohio University, Athens, OH, USA}
\affil[2]{Quantum \& Condensed Matter (T-4) Group, Los Alamos National Laboratory, Los Alamos, NM, USA}
\date{}

\maketitle

\setstretch{1.5}
\setcounter{suppfigure}{1}

\begin{abstract}
A recent work introduced the Simulator Collection for Atomic-to-Continuum Scales (SCACS) toolkit, a framework for improving finite element predictions of heat flow by mapping atom-resolved thermal conductivity into the stiffness matrix of the Galerkin finite element formulation [Ugwumadu \textit{et al.}, Phys. Rev. Materials 10, 053804 (2026)]. Here, we demonstrate that SCACS-derived conductivity fields are solver-independent and can be transferred to existing continuum simulation platforms. As a proof of concept, we map SCACS-derived conductivity fields from complex silicon structures onto finite element meshes in Abaqus\textsuperscript{\textregistered} and compare the resulting heat-flow solutions with those obtained using the native SCACS solver. The agreement between the two implementations shows that atomistically informed conductivity fields can be incorporated into existing finite element workflows without loss of accuracy. This work supports broader efforts to improve the predictive capability of continuum simulations for efficient materials design and property prediction.
\end{abstract}

\section{Data availability and description}\label{sec:aux_data}
The SCACS data-point files and the converted Abaqus input files for the models investigated in this work are provided in Ref. \cite{Zenodo}. The description of the files are provided in Table \ref{tab:SCACS_files} and Table \ref{tab:Abaqus_files} for the SCACS and Abaqus data, respectively. Although the files can be opened using any text editor, visualization, rendering, and post-processing of the SCACS files are best implemented in Paraview \cite{ParaView}, while the Abaqus models require the Abaqus\textsuperscript{\textregistered} computer-aided engineering (CAE) software.

\begin{table}[!tpbh]
\refstepcounter{supptable}
\small
\setlength{\tabcolsep}{6pt}
\renewcommand{\arraystretch}{1.2}
\centering
\caption{Description of SCACS data-point files}
\begin{tabular}{ll}
\hline
\textbf{\texttt{SCACS}}\,/\,Filename & Description\\
\hline
\texttt{M1.xdmf(.h5)}       & twin-grain-boundary Si nanowire  \\
\texttt{M2.xdmf(.h5)}       & amorphous–crystalline Si interface  \\
\texttt{M3.xdmf(.h5)}      & Si nanopillar structure (4 pillars)   \\
\hline
\end{tabular}
\label{tab:SCACS_files}

\end{table}

\begin{table}[!h]
\refstepcounter{supptable}
\small
\setlength{\tabcolsep}{12pt}
\renewcommand{\arraystretch}{1.5}
\centering
\caption{Description of Abaqus input files}
\label{tab:Abaqus_files}
\begin{tabular}{c l c c r r}
\toprule
Model & \textbf{\texttt{ABAQUS}}\,/\,Filename & Solution Type & \makecell{Conductivity} & Elements & Nodes \\
\midrule
\multirow{4}{*}{$M1$} & \texttt{M1\_Dirichlet\_SS\_Aniso.inp}         & Steady State & Anisotropic      & \multirow{4}{*}{1,682,384} & \multirow{4}{*}{162,000}   \\
 &\texttt{M1\_Dirichlet\_Transient\_Aniso.inp}  & Transient    & Anisotropic       &  &  \\
 &\texttt{M1\_Dirichlet\_SS\_Iso.inp}             & Steady State & Isotropic  &  &  \\
 &\texttt{M1\_Dirichlet\_Transient\_Iso.inp}   & Transient    & Isotropic  &  &  \\
\hline 
\multirow{4}{*}{$M2$} &\texttt{M2\_Dirichlet\_SS\_Aniso.inp}                & Steady State & Anisotropic       & \multirow{4}{*}{5,417,280} & \multirow{4}{*}{511,584}   \\
 &\texttt{M2\_Dirichlet\_Transient\_Aniso.inp}      & Transient    & Anisotropic       & &  \\
 &\texttt{M2\_Dirichlet\_SS\_Iso.inp}              & Steady State & Isotropic  & &  \\
 &\texttt{M2\_Dirichlet\_Transient\_Iso.inp}       & Transient    & Isotropic  & &  \\
\hline
\multirow{4}{*}{$M3$} &\texttt{M3\_Dirichlet\_SS\_Aniso.inp}                & Steady State & Anisotropic      & \multirow{4}{*}{462,372} & \multirow{4}{*}{504,621}   \\
 &\texttt{M3\_Dirichlet\_Transient\_Aniso.inp}         & Transient    & Anisotropic      & &  \\
 &\texttt{M3\_Dirichlet\_SS\_Iso.inp}                 & Steady State & Isotropic & &  \\
 &\texttt{M3\_Dirichlet\_Transient\_Iso.inp}       & Transient    & Isotropic & &  \\
\bottomrule
\end{tabular}
\end{table}

\subsection{ParaView-to-Abaqus Data-Conversion Utility}
\label{sec:data_conversion_utility}

A Python utility, \texttt{Paraview\_to\_Abaqus.py}, provided in the \textbf{\texttt{SCACS/}} folder, converts SCACS point-data exported from ParaView (Table \ref{tab:SCACS_files}) into  the Abaqus input files (Table \ref{tab:Abaqus_files}). The utility accepts data in comma-separated value (\texttt{.csv}) or text (\texttt{.txt}) format and automatically identifies comma-, tab-, or whitespace-delimited input. The script also supports visualization of the processed nodal fields using PyVista and conversion of an associated Visualization Toolkit unstructured-grid file (\texttt{.vtu}) to the Abaqus input-file format (\texttt{.inp}) using MeshIO.

The SCACS input file contain the nodal identifiers, spatial coordinates, and components of the thermal-conductivity field. The ParaView column names are mapped to the simplified notation
\begin{equation}
    \left\{
    \texttt{Point ID},\,\texttt{\_0},\,\texttt{\_1},\,\texttt{\_2}
    \right\}
    \longrightarrow
    \left\{
    \texttt{ID},\,\texttt{X},\,\texttt{Y},\,\texttt{Z}
    \right\},
\end{equation}
where \texttt{X}, \texttt{Y}, and \texttt{Z} denote the nodal coordinates. Similarly, the exported conductivity fields
\begin{equation}
    \left\{
    \texttt{k\_xx\_used},\,
    \texttt{k\_yy\_used},\,
    \texttt{k\_zz\_used}
    \right\}
    \longrightarrow
    \left\{
    \texttt{k\_xx},\,
    \texttt{k\_yy},\,
    \texttt{k\_zz}
    \right\}
\end{equation}
are interpreted as the diagonal components $k_{xx}$, $k_{yy}$, and $k_{zz}$ of the thermal-conductivity tensor. An optional scalar conductivity field, denoted by \texttt{k\_used}, that indicate the conductivity of the loading may also be included.

\section{Description of the notations}
Table~\ref{tab:thermal_conductivity_notation} summarizes the notation and corresponding dimensions for the thermal-conductivity fields and finite-element quantities used in the main text. 

\begin{table*}[!tbph]
\refstepcounter{supptable}
\small
\renewcommand{\arraystretch}{2}
\caption{Notation for thermal-conductivity quantities and finite-element matrices.}
\label{tab:thermal_conductivity_notation}
\centering
\begin{threeparttable}
\begin{tabular*}{\linewidth}{@{\extracolsep{\fill}}lll}
\toprule
Quantity
& Notation
& Dimension \\
\midrule

Temperature field\tnote{a}
&
\(T(\overrightarrow{x},t)\)
&
Scalar field \\

Temperature DOF vector\tnote{b}
&
\(\displaystyle
\overrightarrow{T} =
\begin{bmatrix}
T_1 & T_2 & \cdots & T_{N_n}
\end{bmatrix}^{T}
\)
&
\(N_n \times 1\) vector \\

Temperature gradient
&
\(\displaystyle
\nabla T =
\begin{bmatrix}
\frac{\partial T}{\partial x} &
\frac{\partial T}{\partial y} &
\frac{\partial T}{\partial z}
\end{bmatrix}^{T}
\)
&
\(3 \times 1\) vector \\

Heat-flux vector
&
\(\displaystyle
\overrightarrow{q} =
\begin{bmatrix}
q_x & q_y & q_z
\end{bmatrix}^{T}
\)
&
\(3 \times 1\) vector \\

Element conductivity tensor\tnote{c}
&
\(\displaystyle
\overleftrightarrow{K}^{(e)} =
\begin{bmatrix}
K_{xx}^{(e)} & K_{xy}^{(e)} & K_{xz}^{(e)} \\
K_{yx}^{(e)} & K_{yy}^{(e)} & K_{yz}^{(e)} \\
K_{zx}^{(e)} & K_{zy}^{(e)} & K_{zz}^{(e)}
\end{bmatrix}
\)
&
\(3 \times 3\) tensor \\

Elementwise conductivity components
&
\(\displaystyle
K_{xx}^{(e)},\; K_{yy}^{(e)},\; K_{zz}^{(e)}
\)
&
Scalar values \\

Elementwise \(K_{xx}\) field\tnote{d}
&
\(\displaystyle
\overrightarrow{K}_{xx}
=
\left\{K_{xx}^{(e)}\right\}_{e=1}^{n_e}
=
\begin{bmatrix}
K_{xx}^{(1)} &
\cdots &
K_{xx}^{(n_e)}
\end{bmatrix}^{T}
\)
&
\(n_e \times 1\) vector \\

Element FE matrix\tnote{e}
&
\(\overleftrightarrow{K}_{\mathrm{FE}}^{e}\)
&
\(n_e \times n_e\) matrix \\

Element FE matrix entry\tnote{f}
&
\(\displaystyle
K_{ij}^{e}
=
\int_{\Omega_e}
\left(\nabla N_i\right)^{T}
\overleftrightarrow{K}^{(e)}
\nabla N_j
\, d\Omega
\)
&
Scalar \\

Global FE matrix\tnote{g}
&
\(\displaystyle
\overleftrightarrow{K}_{\mathrm{FE}}
=
\mathcal{A}_{e=1}^{n_e}
\left[
\overleftrightarrow{K}_{\mathrm{FE}}^{e}
\right]\)
&
\(N_n \times N_n\) matrix \\

Discretized FE system
&
\(\overleftrightarrow{K}_{\mathrm{FE}}\overrightarrow{T} = \overrightarrow{F}\)
&
Matrix-vector equation \\

Load/source vector
&
\(\displaystyle
\overrightarrow{F} =
\begin{bmatrix}
F_1 & F_2 & \cdots & F_{N_n}
\end{bmatrix}^{T}
\)
&
\(N_n \times 1\) vector \\

\bottomrule
\end{tabular*}

\begin{tablenotes}
\footnotesize
\item[a] Temperature at spatial position \(\overrightarrow{x}\) and time \(t\).
\item[b] Vector of nodal temperature degrees of freedom, where \(N_n\) is the number of global nodes or temperature unknowns.
\item[c] Conductivity tensor assigned to finite element \(e\). This is the mapped per-element SCACS-derived tensor.
\item[d] Collection of the \(K_{11}\) values over all mesh elements, where \(n_e\) is the total number of finite elements.
\item[e] Element-level FE conductivity, or stiffness, matrix obtained from the element conductivity tensor and basis-function gradients.
\item[f] Scalar entry of the element FE matrix coupling local basis functions \(N_i\) and \(N_j\) over element domain \(\Omega_e\).
\item[g] Assembled global finite-element conductivity, or stiffness, matrix. Here, \(\mathcal{A}_{e=1}^{n_e}[\cdot]\) denotes assembly over all \(n_e\) finite elements.
\end{tablenotes}
\end{threeparttable}
\end{table*}

\newpage
\section{Supporting Figures}
\begin{figure*}[!h]
    \centering
    \includegraphics[width=\linewidth]{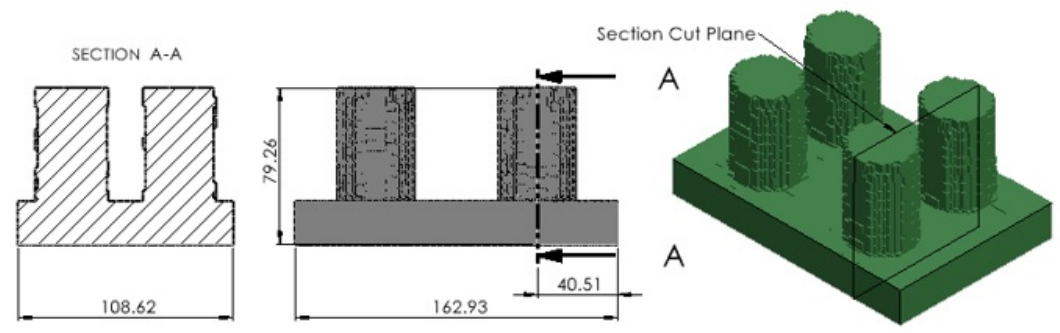}
    \caption{Cross-section of the slice used for $M3$ in this work.}
    \label{fig:Sfig0_Slices}
    \refstepcounter{suppfigure}
\end{figure*}

\begin{figure*}[!h]
    \centering
    \includegraphics[width=\linewidth]{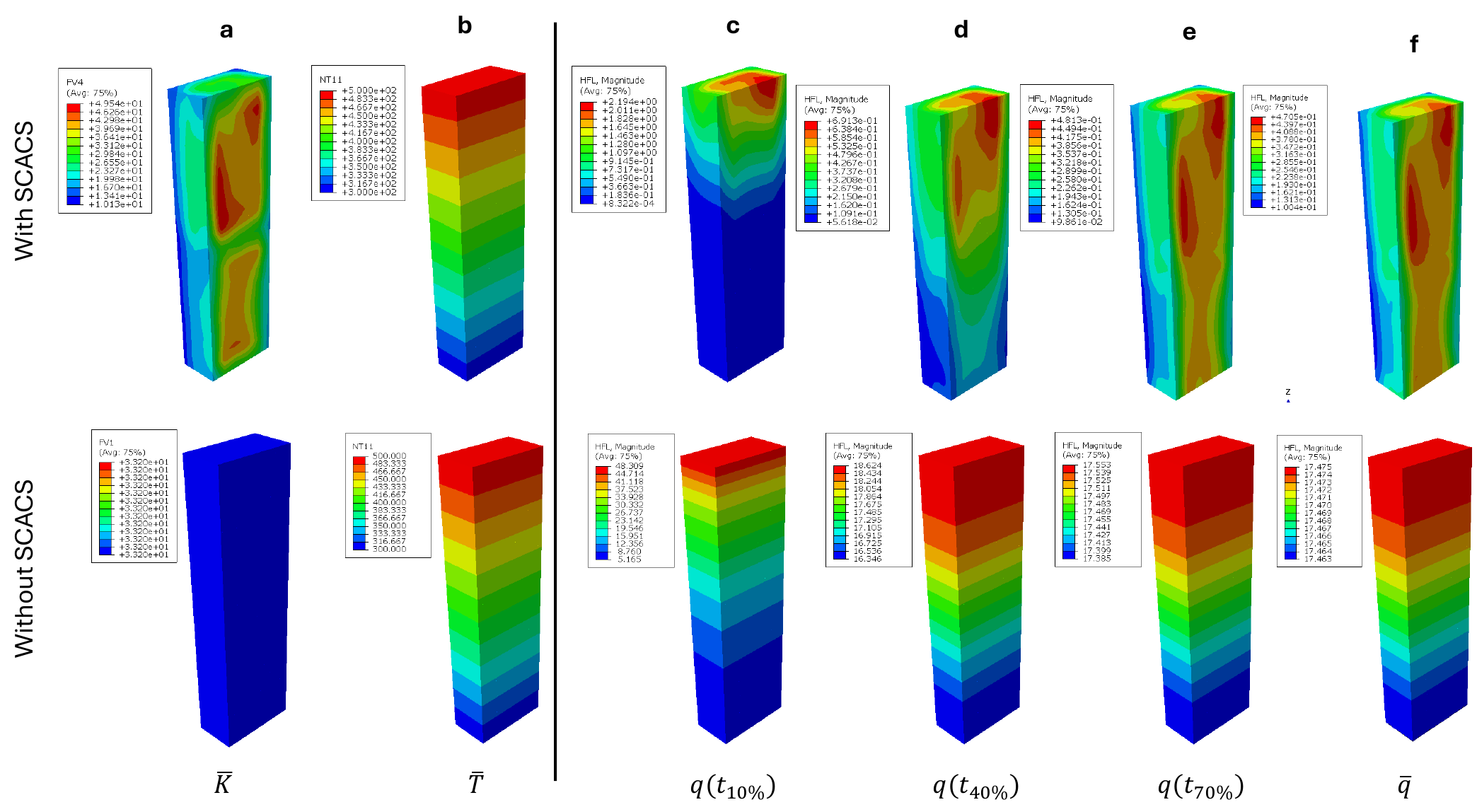}
    \caption{Analysis of the grain-boundary Si nanowire \(M1\). The colormaps show the values of (a) the isotropic conductivity field, $\bar K$, and (b) the steady-state temperature field, $\bar T$. Panels (c--e) show the transient heat-flux, $q_t$, at different times: (c) $t = 10\%$, (d) $t = 40\%$, and (e) $t = 70\%$ of the convergence time. The steady-state heat-flux, $\bar q$, is shown in (f).}
    \label{fig:Sfig1_M1}
    \refstepcounter{suppfigure}
\end{figure*}

\begin{figure*}[!tbhp]
    \centering
    \includegraphics[width=\linewidth]{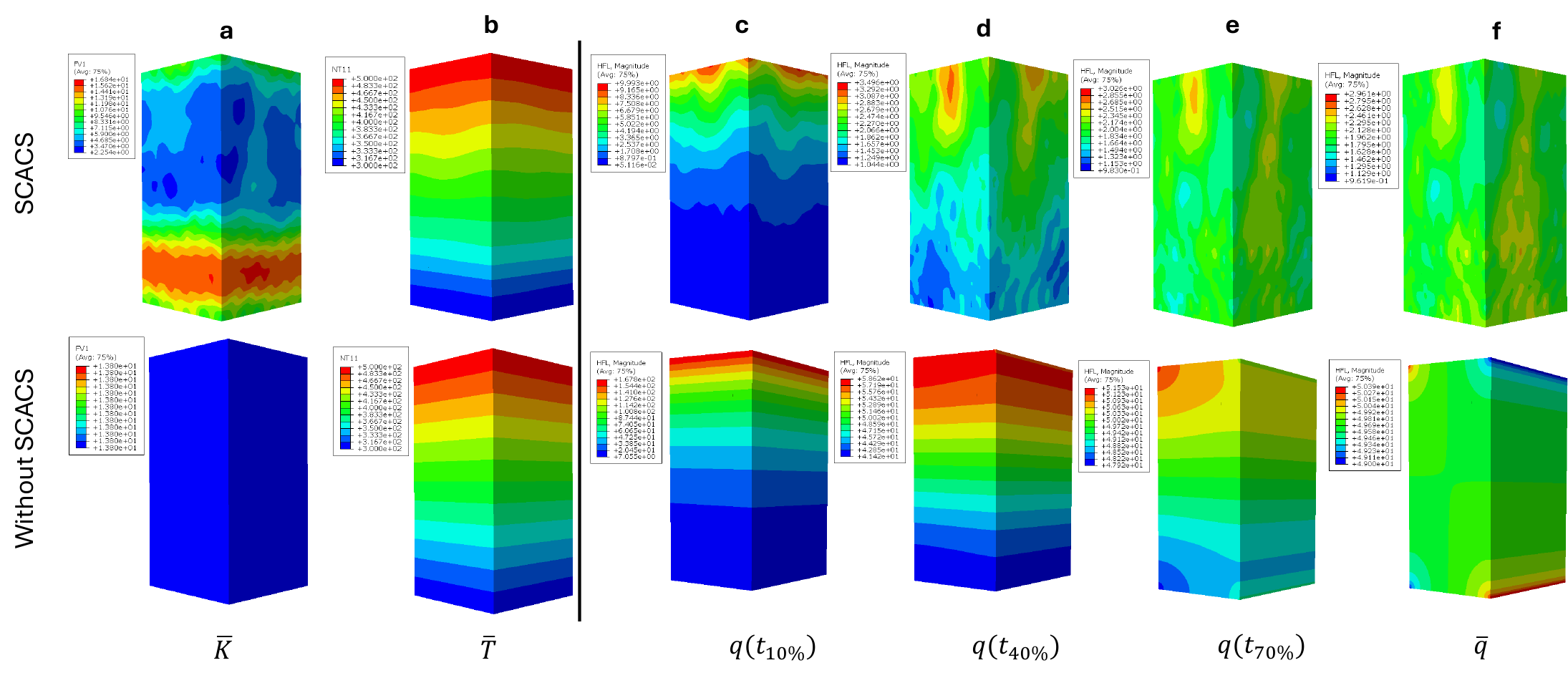}
    \caption{Analysis of the amorphous-crystalline Si interface structure (M2). The colormaps show the values of (a) the isotropic conductivity field, $\bar K$, and (b) the steady-state temperature field, $\bar T$. Panels (c--e) show the transient heat-flux, $q_t$, at different times: (c) $t = 10\%$, (d) $t = 40\%$, and (e) $t = 70\%$ of the convergence time. The steady-state heat-flux, $\bar q$, is shown in (f).}
    \label{fig:Sfig2_M2}
    \refstepcounter{suppfigure}
\end{figure*}

\begin{figure*}[!t]
    \centering
    \includegraphics[width=\linewidth]{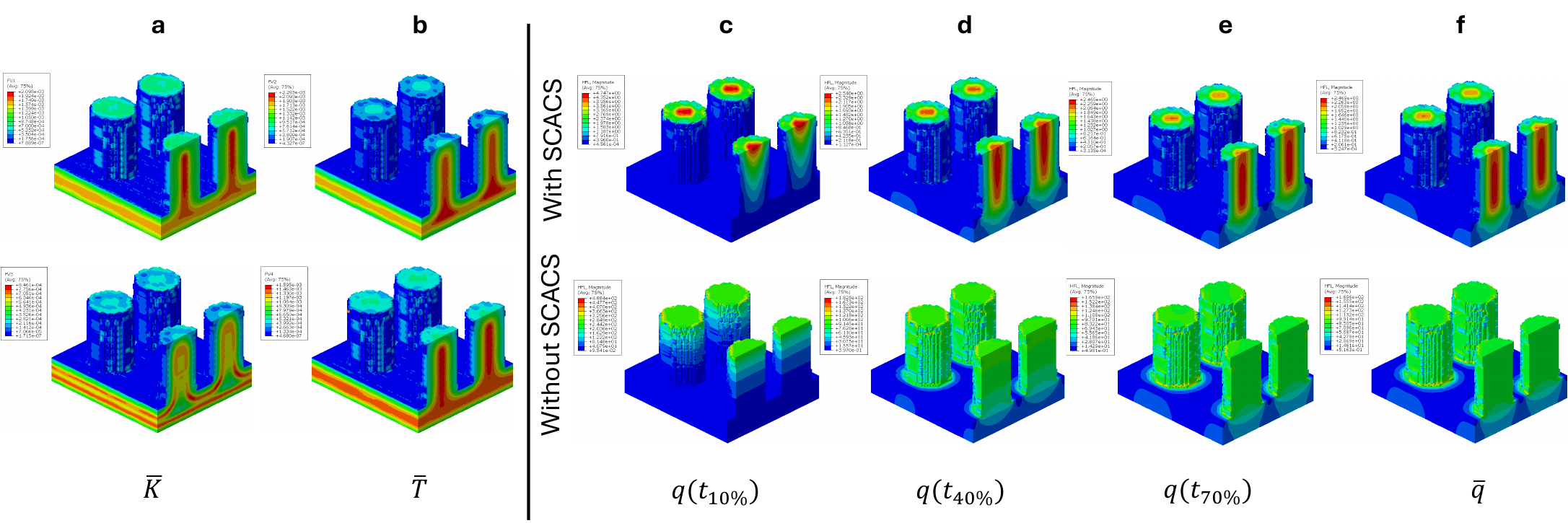}
    \caption{Analysis of the Si nanopillar structure (\(M3\)). Panel (a) shows the elementwise directional conductivity fields, \(K_{xx}^{(e)}\), \(K_{yy}^{(e)}\), and \(K_{zz}^{(e)}\), together with the corresponding isotropic conductivity field, \(\bar K^{(e)}\). Panels (b--d) show the transient heat-flux field, \(q_t\), at (b) \(t=10\%\), (c) \(t=40\%\), and (d) \(t=70\%\) of the convergence time. Panel (e) shows the steady-state heat-flux field, \(\bar q\). The upper and lower rows correspond to simulations performed with and without the SCACS-derived conductivity field, respectively.}
    \label{fig:Sfig3_M3}
    \refstepcounter{suppfigure}
\end{figure*}

Figure \ref{fig:Sfig0_Slices} shows the cross-section for slice considered for $M3$. Figures~\ref{fig:Sfig1_M1}, \ref{fig:Sfig2_M2}, and \ref{fig:Sfig3_M3} reproduce the analyses of \(M1\), \(M2\), and \(M3\) presented in the main text. Here, the colormaps show the actual field values rather than the normalized values used in the main figures. The qualitative trends and physical interpretation remain unchanged.

\newpage

\section*{Supplemental references}
\bibliography{references}